\documentclass[aps,prl,twocolumn,showpacs,superscriptaddress,letterpaper,amsmath,amssymb]{revtex4}

\usepackage[T1]{fontenc} 
\usepackage{graphicx}
\usepackage{graphics}
\usepackage[mathscr]{eucal}
\usepackage{amssymb}
\usepackage{amsmath}
\usepackage{xspace}
\usepackage{listings}
\usepackage{ulem}
\usepackage{color}
\usepackage{ulem}
\usepackage{bm}         
\usepackage{array}
\usepackage{multirow}   
\usepackage{booktabs}   
\usepackage{mleftright}
\usepackage[utf8]{inputenc}
\newcommand*{\Scale}[2][4]{\scalebox{#1}{$#2$}}%

\begin{document}

\title{Exotic Trajectories Effects on Neutrino Oscillations}

\author{Jonathan Miller}
 \affiliation{
Departamento de F\'{\i}sica, Universidad T\'ecnica Federico Santa
Mar\'{\i}a\\ Casilla 110-V, Valpara\'iso, Chile;\\ e-mail: Jonathan.Miller@usm.cl}

\author{Roman Pasechnik}
 \affiliation{
Theoretical High Energy Physics, Department of Astronomy and
Theoretical Physics, Lund University, S\"olvegatan 14A, SE 223-62
Lund, Sweden;\\ e-mail: Roman.Pasechnik@thep.lu.se}

\begin{abstract}
Recently, for the first time, exotic loop trajectories were observed in a photon triple slit experiment. We discuss possible origins and potential impacts of analogical exotic trajectories in neutrino production, 
propagation and detection onto the neutrino oscillation observables.
\end{abstract}


\maketitle

\noindent \textbf{\textit{Introduction.}}~A neutrino interacts with matter via the electroweak force in the flavor basis while it inherently propagates 
as a mass (or matter) eigenstate \cite{Blennow:2013rca}. This allows for an abundance of interesting phenomena, many of which are still active topics 
of research. For a comprehensive review on the current status of this research field, see e.g.~Refs.~\cite{Bellini:2013wra,Maltoni:2015kca,Beuthe:2001rc}.

In the Quantum Field Theory (QFT) approach the neutrino production, propagation, and detection processes are considered as the single neutrino 
system~\cite{Akhmedov:2010ms}. In certain cases, due to the physical neutrino mass and mixing parameters or the relevant formalism, the neutrino 
system can be viewed as a two-state system. One may consider a two-state neutrino of definite energy as being split to two separate space-time paths 
upon creation which develop a phase over time in the course of its propagation. 
\begin{figure}[!h]
 \centerline{\includegraphics[width=0.3\textwidth]{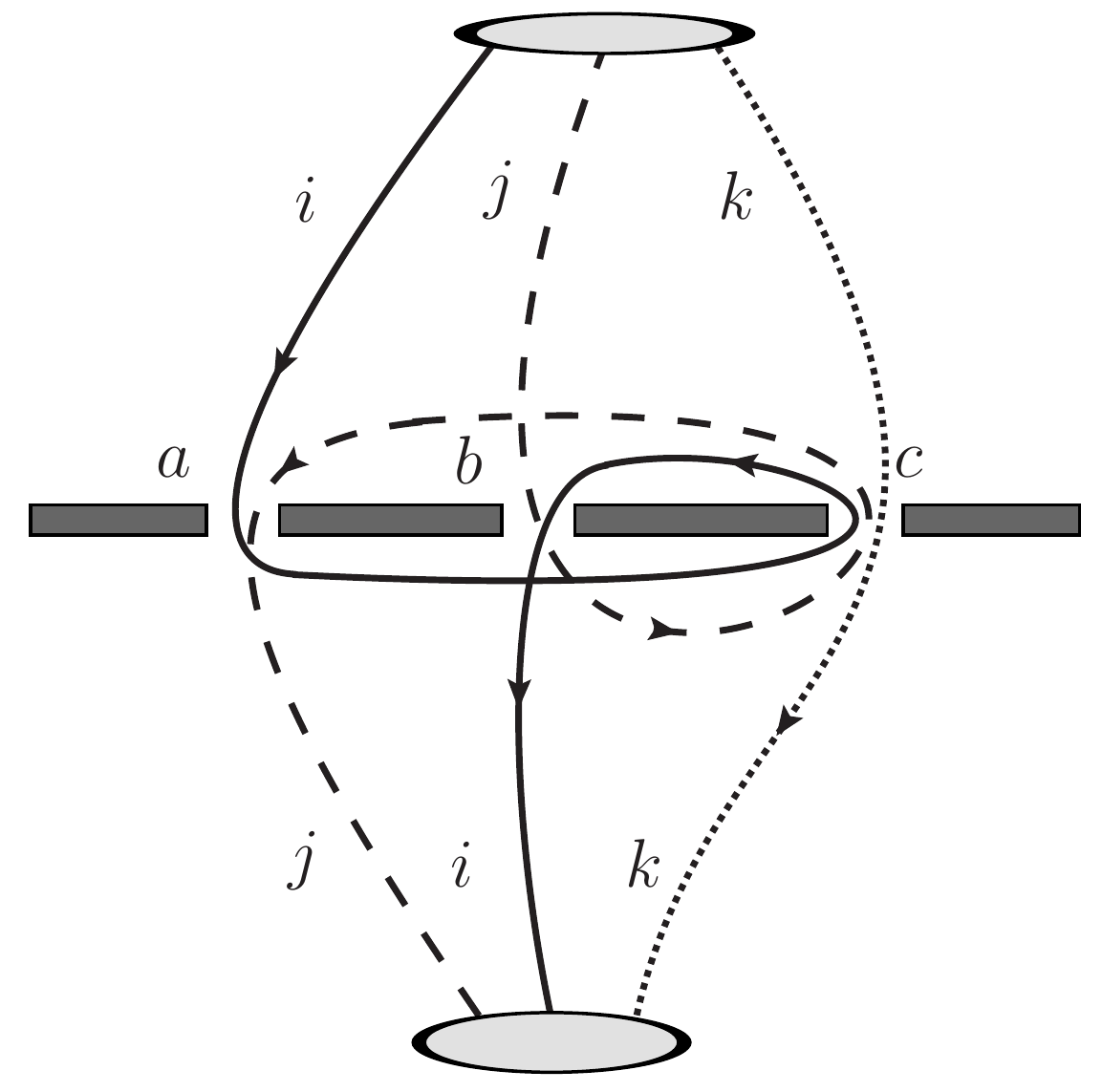}}
   \caption{ \small 
   Exotic (non-classical) loop trajectories of photons in a triple-slit measurement illustrating a possible 
   ``switch'' between $i$ and $j\not=i$ paths in the presence of the third slit $c$.}
 \label{fig:triple-slit}
\end{figure}

Recently, so-called exotic loop trajectories have been detected in photon triple slit experiments \cite{Magana-Loaiza:2016}. 
The trajectories, and the observed associated interference patterns, depend on the structure of the slits and the excitation of the slit near field. 
These new exotic trajectories are not expected in a Quantum Mechanical formulation but naturally emerge in the QFT framework 
and are apparent in its Path Integral formulation. A relevant example of the photon exotic loop trajectories relevant for triple slit interferometry is 
shown in Fig.~\ref{fig:triple-slit}. These new exotic trajectories have been described by including an additional propagator between 
the slits (see Ref.~\cite{Sawant:2013cea}) which causes an additional phase difference to accrue.
\begin{figure}[!h]
 \centerline{\includegraphics[width=0.3\textwidth]{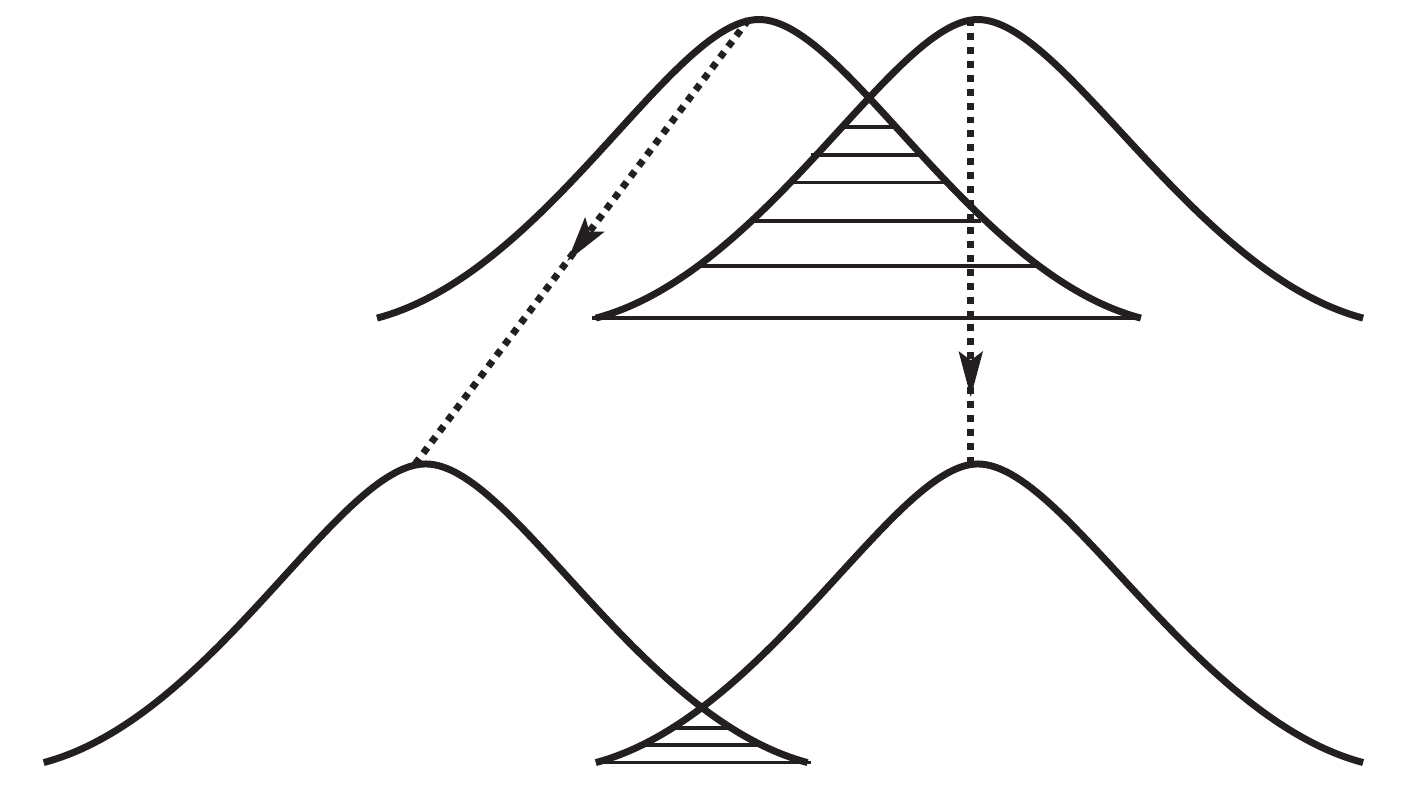}}
   \caption{ \small 
   An illustration of the time evolution of the ovelapping neutrino wave packets 
   $\nu_1(t)$ and $\nu_2(t)$ in the course of their propagation.}
 \label{fig:packets}
\end{figure}

The neutrino mass states in a superposition also propagate over the separate space-time paths representing an analogy with the slits in a photon 
interferometry experiment. Thus similar neutrino exotic trajectories may exist in neutrino interference measurements. The existence of an overlap between 
the propagating mass states (see Fig.~\ref{fig:packets}) provides an opportunity for a virtual particle to travel between the mass eigenstates, accruing a phase shift, 
and resulting in a small amplitude for $\nu_i$ and $\nu_j$ to switch their paths. In this Letter, we discuss this phenomena in the three flavor case in more 
detail, when it might be relevant and how the standard description of neutrino oscillations, derived from ``classical'' trajectories, may be modified if 
the neutrino exotic trajectories were relevant.

\noindent \textbf{\textit{Exotic neutrino trajectories.}}~Exotic trajectories in neutrino oscillation are due to the three-state nature of neutrinos and would not 
be present if neutrinos only existed in two states. Similar to the triple-slit experiments, the neutrino mass state propagates to some point $x$ along the neutrino 
path where a ``switch'' occurs and the phase continues to accrue with a small phase shift, $\delta_{ij}$, for the trajectories $\theta_{i}$ and $\theta_{j}$ 
but not for trajectory $\theta_{k}$. This switch is, in particular, possible due to an overlap between eigenstates $\nu_i$ and $\nu_j$.

\begin{center}
\begin{figure}[!h]
\includegraphics[width=0.7\textwidth]{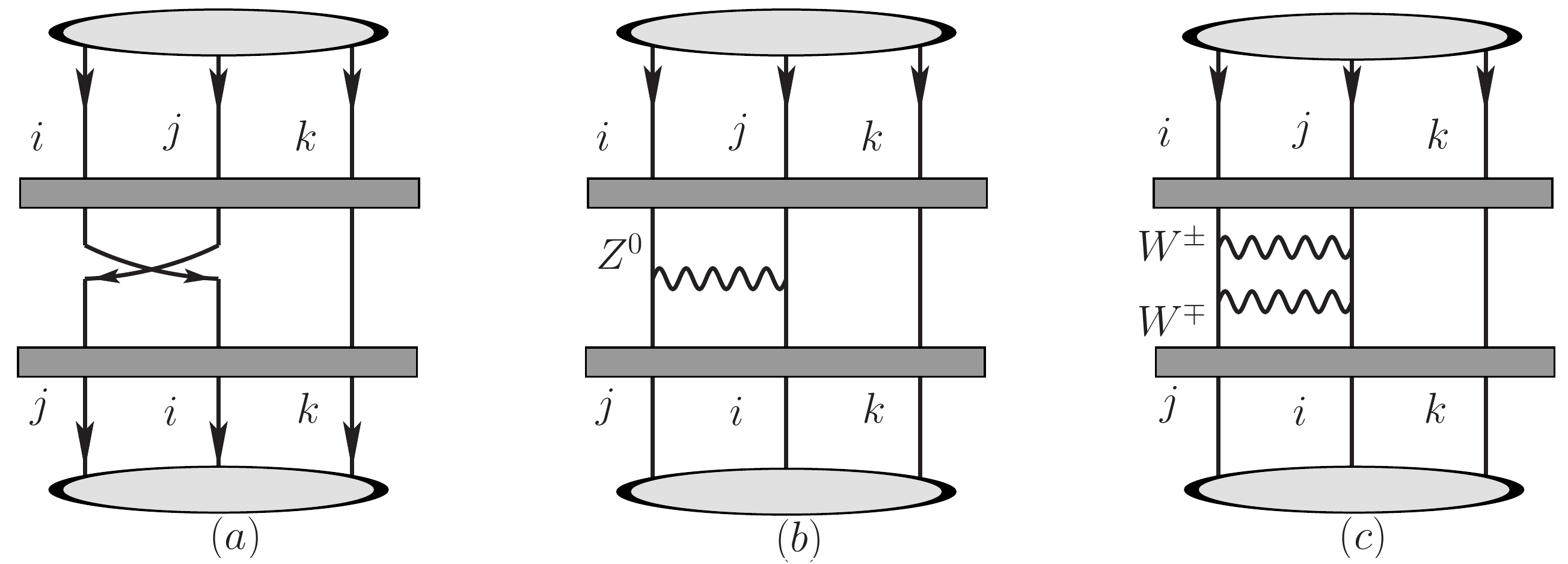}
   \caption{ \small 
   Possible mechanisms causing a ``switch'' of spacetime paths of the neutrino mass states $i$ and $j\not=i$  in the course 
   of their propagation in the three-flavor case ($i,j,k=1,2,3$): (a) virtual transverse motion due to an overlap of wave packets 
   of the $\nu_i$ and $\nu_j$ mass states, (b) non-standard (flavour-changing) neutral-current interactions via a virtual $Z^0$ 
   boson exchange between the neutrino trajectories, (c) charged-current interactions via a virtual $W^\pm$ pair exchange between 
   the neutrino trajectories. The presence of a third mass state $k\not=i,j$ is relevant for acummulation of non-trivial phase differences 
   affecting the three-flavor neutrino oscillation observables.}
 \label{fig:switch}
\end{figure}
\end{center}

Since the additional phase for trajectories $\theta_{i}$ and $\theta_{j}$ is the same, the phase difference between these trajectories displays 
the standard behavior. However, the phase differences $\theta_{ik}$ and $\theta_{jk}$ will acquire an additional phase $\delta_{ij}$.
The amplitude for these exotic phase trajectories, like those for the photon, are small. We have identified three different processes which 
may allow this ``switch''. One is the process see in Fig.~\ref{fig:switch}b where a $Z$ boson is exchanged between the two propagating 
mass states. While this is not a process found in the Standard Model, it is a process considered by physicists searching for the so-called Non Standard 
Interactions where an exchange of a $Z$ can change the neutrino state \cite{Miranda:2015dra}. This trajectory would be suppressed 
by a factor of $g^2$. Another such process, see Fig.~\ref{fig:switch}c, is the box process where two $W$ bosons are exchanged between the two 
neutrino eigenstates. This trajectory would be suppressed by a loop factor and $g^4$.
\onecolumngrid
\twocolumngrid

Finally, we can consider virtual neutrino propagators between the two neutrino states in energy/momentum space as seen in Fig.~\ref{fig:switch}a. 
This provides us with little information about the processes involved and the amplitude of the exotic trajectory but does give us a first estimate 
for the additional accrued phase $\delta_{ij}$ and displays some of the features of such exotic trajectories in generic terms. It is this process which 
most closely mirrors the three slit (see Fig.~\ref{fig:triple-slit}) exotic trajectories studied and observed in Refs.~\cite{Sawant:2013cea} and 
\cite{Magana-Loaiza:2016}.

The amplitude or existence of the exotic photon trajectories in Ref.~\cite{Magana-Loaiza:2016} depend crucially on the size of the photon wave function. 
Similarly, the processes which allow the exotic trajectories in neutrino oscillation all depend on the overlap (and thus size, $\sigma_E$) of the neutrino 
wave packet \cite{Beuthe:2001rc}. Explicitly, both the Beyond the Standard Model process in Fig.~\ref{fig:switch}b and the Standard Model process in 
Fig.~\ref{fig:switch}c require the neutrino states to be close. Additionally, the spacing of the slits matters in Ref.~\cite{Sawant:2013cea} and 
\cite{Magana-Loaiza:2016} and similarly the amplitude of the interference due to the exotic trajectories will depend on the inverse of the mass difference. 
To summarize, the amplitude for a switch at $t_x$ behaves as
\begin{equation}
A^{e}_{ij}\left(t\right) \propto \frac{M_{ij}\left(\sigma_{E},E,t_x\right)}{\Delta m_{ij}} \,,
\end{equation} 
where $E$ is the neutrino energy, $M_{ij}$ is the overlap function between $\nu_i$ and $\nu_j$, and $\Delta m_{ij}=m_i-m_j$.

Using the traditional evolution language of neutrino oscillations we can describe the evolution of a flavor state by considering a neutrino produced in 
a flavor state $\alpha=e,\,\mu,\,\tau$ at $t=0$ as a superposition of three mass states $|\nu_\alpha(0)\rangle = \sum_i V_{\alpha i}^* |\nu_i(0) \rangle.$
Here, $V_{\alpha i}$ are the traditional PMNS matrix elements
\begin{multline*}
V = 
\Scale[0.85]{\begin{pmatrix} 
c_{12}c_{13} & s_{12}c_{13}  & s_{13}e^{- i \delta_{\rm cp}} \\
-s_{12}c_{23}-c_{12}s_{13}s_{23}e^{i \delta}  & c_{12}c_{23}-s_{12}s_{13}s_{23}e^{i \delta_{\rm cp}}  & s_{23}c_{13}  \\
s_{12}s_{23}-c_{12}s_{13}c_{23}e^{i \delta}  & -c_{12}s_{23}-s_{12}s_{13}c_{23}e^{i \delta_{\rm cp}}  & c_{23}c_{13}  
 \end{pmatrix}\,,}
\end{multline*}
where $\delta_{\rm cp}$ is the charge-parity violating phase, $c_{ij}=\cos\theta_{ij}$ and  $s_{ij}=\sin\theta_{ij}$.
The three mass states propagate along three separate spacetime paths and display interference phenomena analogously to triple-slit photon experiments. 
In traditional quantum-mechanical (wave-packet) treatment of neutrino oscillations \cite{Akhmedov:2012mk}, the time evolution of a given mass state 
is completely independent on evolution of other mass states. However, this is not the case for neutrinos propagating through matter with sharp density 
profiles \cite{Blennow:2013rca}. In the case of a sharp density profile from vacuum to the Earth, the flavor from all (overlapping) mass states at the boundary 
defines the matter states which continue to propagate through the Earth (see Ref.~\cite{Miller:2013wta} for explicit two-flavor calculation and for an 
analysis of two-flavor non-overlapping mass states).

The propagating state then picks up a phase depending on neutrino energy $E_i$ and mass $m_{i}$
\begin{eqnarray}
|\nu_\alpha(t)\rangle = \sum_{\beta=e,\,\mu,\,\tau}\Big[\sum_i V_{\alpha i}^* e^{-iE_i t} 
V_{\beta i}\Big] |\nu_\beta(0)\rangle
\end{eqnarray}
giving rise to the standard probability amplitude of $\nu_\alpha \to \nu_\beta$ transition at the time of detection $t=L$ ,
\begin{eqnarray*}
P_{\alpha\beta}(T)\equiv |\langle \nu_{\beta}(0)|\nu_{\alpha}(T)\rangle|^2 = 
\sum_{i,j} V_{\alpha i}^*V_{\beta i}V_{\alpha j}V_{\beta j}^* e^{-i(E_i-E_j)T} 
\end{eqnarray*}
where $ T \left( E_i-E_j \right) \simeq \Delta m_{ij}^2 L/(2 E)$ for a neutrino of energy $E$ \cite{Beuthe:2001rc}.

Following the lead of Refs.~\cite{Ohlsson:1999um} and \cite{Sawant:2013cea}, we can describe the process in Fig.~\ref{fig:switch}a 
by dividing the neutrino propagation into three periods (production to point $x$ which is at time $t_{x}$, the virtual propagation at point $x$, 
and from point $x$ to the detector at $L$ and time $t_{D}$). The flavor evolution operator describing this flavor oscillation in the vacuum 
can be expressed as
\begin{equation}
U_{f}(L)=U_{f}(L,x)U_{s}U_{f}(x,0) \,,
\end{equation}
where $U_{f}(L,x)$ is the standard propagation of the neutrino mass states from point $x$ to point $L$ and can be calculated as 
$U_f(L,x)=V U_m(L,x) V^{-1}$, where $V$ is the PMNS matrix (for vacuum) and $U_m(L,x)$ is the evolution operator between $L$ 
and $x$ in the mass basis (Refs.~\cite{Ohlsson:1999um,Ohlsson:1999xb} have these generalized to matter).

The switch operator in Fig.~\ref{fig:switch}a, for the case $k=3$, can be described in the mass basis as
\begin{equation}\label{eq:switchoperator}
U_s = 
\begin{pmatrix} 
e^{i\delta_{12}} & 0 & 0 \\
0 & e^{i\delta_{12}} & 0 \\
0 & 0 & 1 
 \end{pmatrix}
\end{equation}
and $\delta_{12}$ is the phase accrued by the exotic trajectory. 

This example is illustrative of features which are common to the $U_s$ operator. First, to leading order the same phase change is observed 
for two of the trajectories, $\theta_i$ and $\theta_j$, while the third is unchanged. For the process in Fig.~\ref{fig:switch}a, this means that 
the relative phase difference between the switched trajectories is unchanged while the phase difference between the unchanged trajectory and 
the switched trajectories is changed. It is this phase difference which is an observable in the interference probability. Second, in the proper basis 
the exotic trajectories can be described as a diagonal matrix. In the special case where the switch operator is diagonal in the mass basis, the propagation 
operators display the property of commutativity and the interference is independent of $x$. Finally, the $U_s$ operator is unitary.

The process of the propagation of a virtual neutrino (see Fig.~\ref{fig:switch}a) obviously follows from the matter ``jump'' discussed in the neutrino 
literature \cite{Blennow:2013rca}. In the formalism that successfully describes this jump in neutrino oscillation physics the propagating matter 
states are projected to the flavor basis on one side of the ``jump'' and then projected to the new matter basis on the other side of the ``jump''. 
In the language of particle physics and QFT this is due to an exchange of a virtual particle between the propagating  states. A similar diagram is 
described in Fig.~\ref{fig:switch}a where the off shell virtual particle explicitly rearranges the phase as described in Eq.~(\ref{eq:switchoperator}).

\noindent \textbf{\textit{Discussion.}}~The exotic trajectories observed by Ref.~\cite{Magana-Loaiza:2016} were not observable without three 
important conditions. The first of these is the condition that the experiment was a triple-slit experiment where three photon paths interfere. 
In neutrinos we also find that three masses or three neutrino paths are required. The second is that the photon wave function is large compared 
to the slit spacing; similarly in the three processes in Fig.~\ref{fig:switch} the amplitude of the process depends directly on the overlap between 
the wave packets. Finally, they excited the metal around the three slits to increase the reach of the virtual photon. Similarly, one expect that in 
the present of a strong magnetic field (or an electromagnetic potential) that the ``box'' amplitude in Fig.~\ref{fig:switch} 
would increase.
\begin{figure}[!h]
 \centerline{\includegraphics[width=0.4\textwidth]{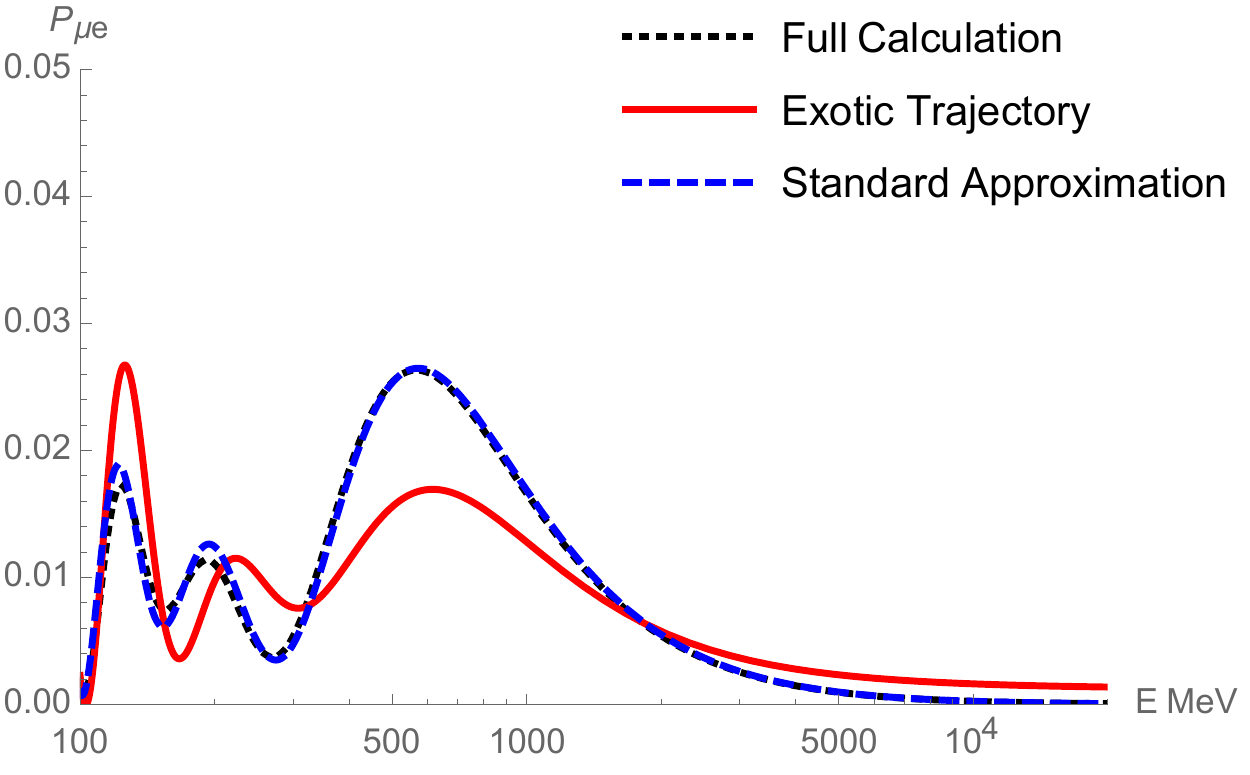}}
   \caption{ \small 
The effect on the vacuum probability at a distance of $L=1300$ km, in the normal mass hierarchy, with mixing coefficients and mass differences 
as in Ref.~\cite{Olive:2016xmw} and $\delta_{\rm cp}=\pi/2$, of ``electron neutrino appearance'' considering the case where there has been 
an exotic phase of $\delta_{13}=0.1$ included. The standard approximation can be found in Ref.~\cite{Nunokawa:2007qh} 
as~$P_{\mu e}=P_{\rm sol}+P_{\rm atm}+2\sqrt{P_{\rm atm}P_{\rm sol}}\cos{\left(\Delta m_{32}^2 L/(4E) + \delta_{\rm cp} \right)}$, 
where~$\sqrt{P_{\rm atm}}=\sin\theta_{23}\sin2\theta_{13}\sin\left(\Delta m_{31}^2 L/(4 E) \right)$~and~$\sqrt{P_{\rm sol}}=\cos\theta_{23}
\cos\theta_{13}\sin2\theta_{12}\sin\left(\Delta m_{21}^2 L/(4 E) \right)$. }
   \label{fig:me13}
\end{figure}
\begin{figure}[!h]
 \centerline{\includegraphics[width=0.4\textwidth]{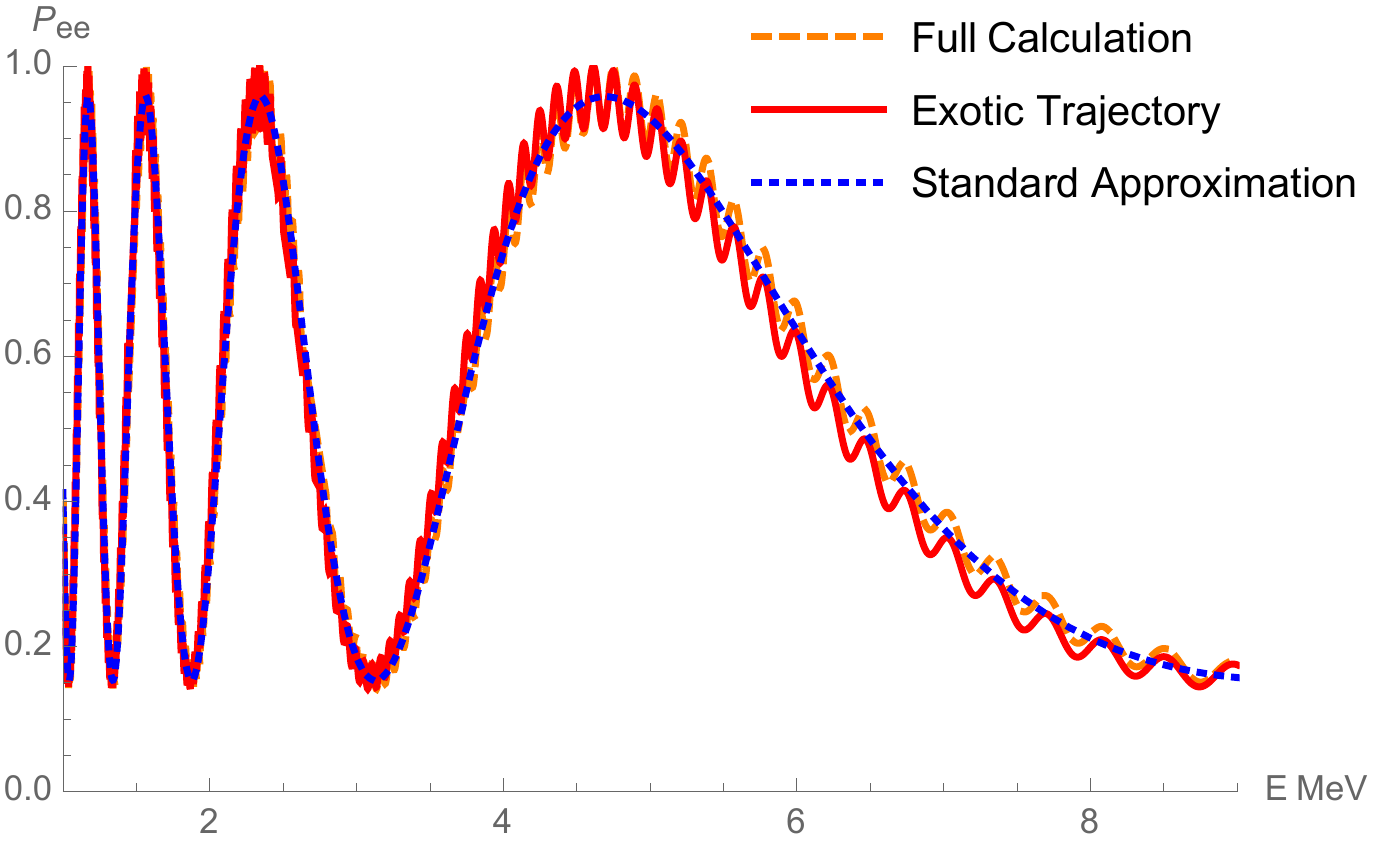}}
   \caption{ \small 
The effect on the vacuum probability at a distance of $L=800$ km, in the normal mass hierarchy, with mixing coefficients and mass differences 
as in Ref.~\cite{Olive:2016xmw} and $\delta_{\rm cp}=\pi/2$, of ``electron neutrino disappearance'' considering the case where there has 
been an exotic phase of $\delta_{13}=0.1$ included. The standard approximation can be found in Ref.~\cite{Gonzalez-Garcia:2014bfa} as 
$P_{ee}=\sin^4\theta_{13}+\cos^4\theta_{13}\left(1-\frac{1}{2}\sin^22\theta_{12}\sin\left(\Delta m_{21}^2 L/(2 E)\right)^2\right)$. }
   \label{fig:ee13}
\end{figure}

Since we are assuming that the overall amplitude is small, we can consider a perturbative expansion in this amplitude. The first term with 
a single 	``switch'' will be the largest term and so the most relevant interference of these exotic trajectories. By looking at the traditional 
probabilities of interest to the neutrino community at the single ``switch'' level we can understand how to observe the impact of such trajectories.

We can estimate the size of the phase change in Fig.~\ref{fig:switch}a caused by a single ``switch'' by looking at the calculation in Ref.~\cite{Sawant:2013cea}. 
We expect the phase accrued by the exotic trajectory to be on the order of
\begin{equation}
\delta_{ij} \simeq \frac{\Delta m_{ij}^2 D_{ij}\left( E, \sigma_E, t_x \right) }{2 E} \,,
\end{equation}
where $D_{ij}\left( E, \sigma_E, t_x \right)$ is the distance between the packets at position $x$. The factor $\Delta m_{ij}^2/E$ makes 
$\delta_{ij}$ very small compared to the phases that accrue due to ``classical'' trajectories. 
The diagrams in Fig.~\ref{fig:switch}b and \ref{fig:switch}c will have additional contributions to the phase. Additionally, $U_s$ for Fig.~\ref{fig:switch}a 
is diagonal in the mass basis and so the phases for large number of ``switches'' may accumulate.

\begin{figure}[!h]
 \centerline{\includegraphics[width=0.4\textwidth]{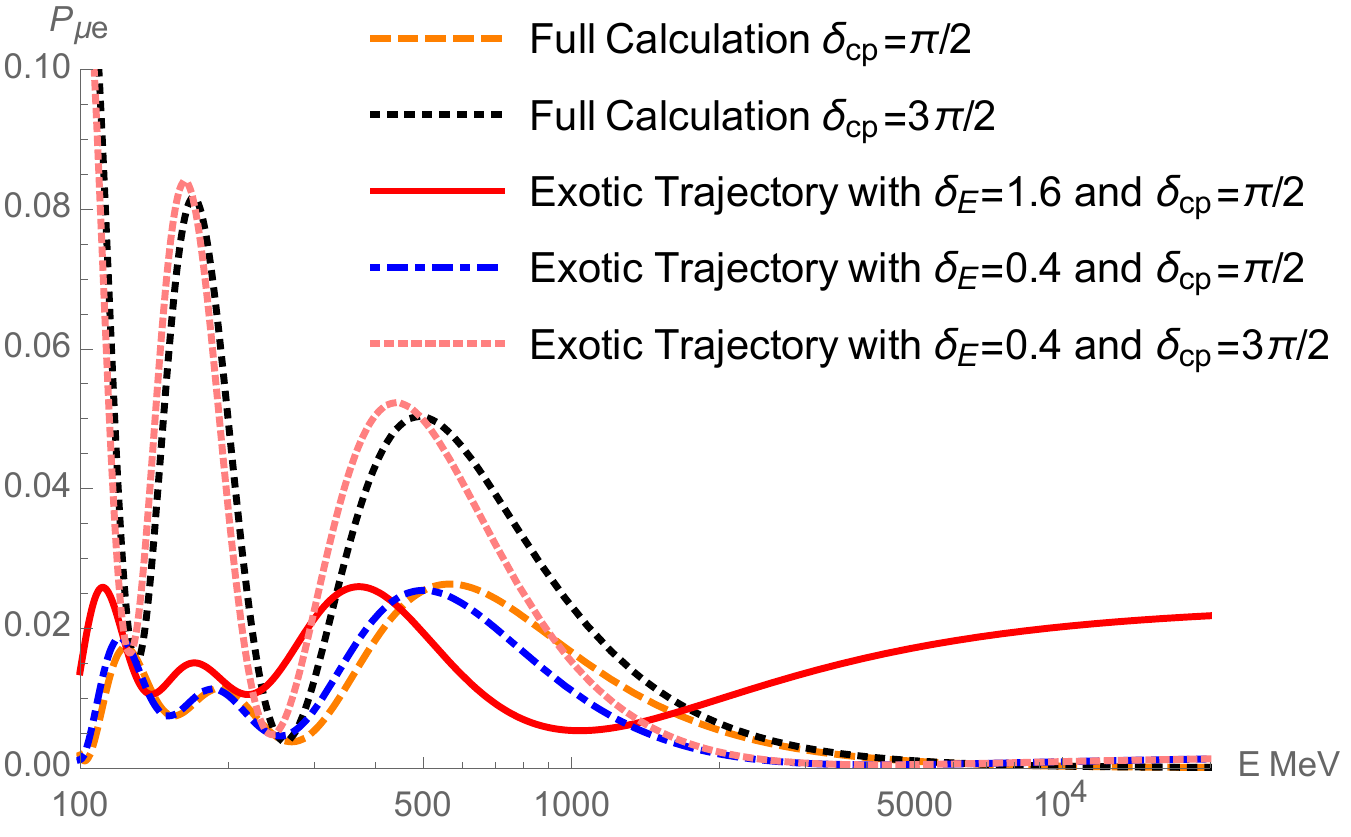}}
   \caption{ \small 
The effect on the vacuum probability at a distance of $L=1300$ km, in the normal mass hierarchy, with mixing coefficients and mass differences 
as in Ref.~\cite{Olive:2016xmw}, of ``electron neutrino appearance'' with representative exotic phases, $\delta_{E}=\delta_{13}=\delta_{23}=\delta_{12}$, 
but with $A^e_{12}=100 A^e_{13}$, $100 A^e_{23}$. }
   \label{fig:mecomplete}
\end{figure}

To understand the impact of a small but observable accrued phase we consider the interference due to an exotic trajectory where the trajectory experiences 
a ``switch'' between the $m_1$ and $m_3$ mass states with an overall phase change of $\delta_{13}=0.1$. In Figs.~\ref{fig:me13} and \ref{fig:ee13}  we show 
the impact of this phase change for ``electron neutrino appearance'' and ``electron neutrino disappearance''. In Fig.~\ref{fig:mecomplete} we consider the change 
in the behavior of ``electron neutrino appearance'' due to larger phase changes. These provide examples of the size and types of changes in the interference pattern 
possible due to the exotic trajectories.

There are some important consequences to consider which we have not studied in detail. One such consequence is on the coherence length. The phase changes from 
the exotic trajectories can increase or decrease the coherence length, the distance from the source of neutrinos to where the wave packets become so separated 
that interference is not observed. For the simple description in Fig.~\ref{fig:switch}a we expect that the coherence length will increase the coherence length. Also, these exotic trajectories may be common when there are collective oscillations such as the early universe or in a supernova.

\par
\noindent \textbf{\textit{Acknowledgments.}}~J.~M.~was supported in part by PROYECTO BASAL FB 0821 CCTVal, by CONICYT grant 
PIA ACT1413 and Fondecyt (Grant No.~11130133). R.~P.~was partially supported by the Swedish Research Council, contract 
number 621-2013-428, by CONICYT grant PIA ACT1406 and Fondecyt (Grant No.~11130133).




\begin{thebibliography}{99}

\bibitem{Blennow:2013rca} 
 M.~Blennow and A.~Y.~Smirnov,
 Adv.\ High Energy Phys.\  {\bf 2013}, 972485 (2013).

\bibitem{Bellini:2013wra} 
  G.~Bellini, L.~Ludhova, G.~Ranucci and F.~L.~Villante,
  Adv.\ High Energy Phys.\  {\bf 2014}, 191960 (2014)

\bibitem{Maltoni:2015kca} 
  M.~Maltoni and A.~Y.~Smirnov,
  Eur.\ Phys.\ J.\ A {\bf 52}, no. 4, 87 (2016)

\bibitem{Beuthe:2001rc} 
  M.~Beuthe,
  Phys.\ Rept.\  {\bf 375}, 105 (2003)

\bibitem{Akhmedov:2010ms} 
  E.~K.~Akhmedov and J.~Kopp,
  JHEP {\bf 1004}, 008 (2010)
  Erratum: [JHEP {\bf 1310}, 052 (2013)].







  
\bibitem{Magana-Loaiza:2016}
O.S.~Maga\~na-Loaiza {\it et al}, 
  Nat. Commun. {\bf 7}, 13987 (2016).

\bibitem{Sawant:2013cea} 
  R.~Sawant {\it et al},
  Phys.\ Rev.\ Lett.\  {\bf 113}, no. 12, 120406 (2014).


\bibitem{Miranda:2015dra} 
  O.~G.~Miranda and H.~Nunokawa,
  New J.\ Phys.\  {\bf 17}, no. 9, 095002 (2015)


\bibitem{Akhmedov:2012mk} 
 E.~K.~Akhmedov and A.~Wilhelm,
 JHEP {\bf 1301}, 165 (2013)

\bibitem{Miller:2013wta} 
  J.~Miller and R.~Pasechnik,
  Adv.\ High Energy Phys.\  {\bf 2015}, 381569 (2015).

\bibitem{Ohlsson:1999um} 
  T.~Ohlsson and H.~Snellman,
  Phys.\ Lett.\ B {\bf 474}, 153 (2000)
  Erratum: [Phys.\ Lett.\ B {\bf 480}, 419 (2000)]
 
\bibitem{Ohlsson:1999xb} 
  T.~Ohlsson and H.~Snellman,
  J.\ Math.\ Phys.\  {\bf 41}, 2768 (2000)
  [Erratum-ibid.\  {\bf 42}, 2345 (2001)].









\bibitem{Olive:2016xmw} 
  C.~Patrignani {\it et al.} [Particle Data Group],
  Chin.\ Phys.\ C {\bf 40}, no. 10, 100001 (2016).




\bibitem{Nunokawa:2007qh} 
  H.~Nunokawa, S.~J.~Parke and J.~W.~F.~Valle,
  Prog.\ Part.\ Nucl.\ Phys.\  {\bf 60}, 338 (2008)


\bibitem{Gonzalez-Garcia:2014bfa} 
  M.~C.~Gonzalez-Garcia, M.~Maltoni and T.~Schwetz,
  JHEP {\bf 1411}, 052 (2014)
\end{thebibliography}
\end{document}